\documentclass{elsart}
\usepackage{epsfig}
\usepackage{natbib}
\usepackage{amssymb}
\usepackage{psfrag}
\usepackage{rotating}

\begin{document}
\newcommand{\be}{\begin{equation}}
\newcommand{\ee}{\end{equation}}
\newcommand{\ba}{\begin{eqnarray}}
\newcommand{\ea}{\end{eqnarray}}
\newcommand{\ol}{\overline}
\newcommand{\grad}{\nabla}
\newcommand{\la}{\langle}
\newcommand{\ra}{\rangle}
\newcommand{\Par}{\parallel}

\begin{frontmatter}

\title{Adaptive Mesh Fluid Simulations on GPU}
\author{Peng Wang,}
\author{Tom Abel,}
\author{Ralf Kaehler}
\address{Kavli Institute for Particle Astrophysics and Cosmology, \\ SLAC National Accelerator Center and Stanford Physics Department, Menlo Park, CA 94025}

\begin{abstract}
We describe an implementation of compressible inviscid fluid solvers
with block-structured adaptive mesh refinement on Graphics
Processing Units using NVIDIA's CUDA. We show that a class of
high resolution shock capturing schemes can be mapped naturally on
this architecture. Using the method of lines approach with the second order total
variation diminishing Runge-Kutta time integration scheme, piecewise linear reconstruction, and 
a Harten-Lax-van Leer Riemann solver, we achieve an overall speedup of approximately 10
times faster execution on one graphics card as compared to a single
core on the host computer.  We attain this speedup in uniform grid
runs as well as in problems with deep AMR hierarchies. Our
framework can readily be applied to more general systems of
conservation laws and extended to higher order shock capturing schemes. This is
shown directly by an implementation of a magneto-hydrodynamic solver
and comparing its performance to the pure hydrodynamic case. Finally, we also combined our CUDA 
parallel scheme with MPI to make the code run on GPU clusters. Close to ideal speedup is observed on up to four GPUs.
\end{abstract}


\end{frontmatter}

\section{Introduction}
\label{sec:introduction}

Graphics Processing Units (GPUs) are specialized for math-intensive highly parallel computation, thus more transistors are
devoted to data processing rather than data caching and flow control like in CPU. So the potential
tremendous performance of general non-graphics
computations on GPUs has recently motivated a lot of research activities on general-purpose GPU (GPGPU) computing (see. e.g. Owens et al. 2007 for a review).

NVIDIA introduced the Tesla unified graphics and computing architecture in November 2006. The Tesla architecture is built around
a scalable array of multithreaded streaming multiprocessors (SM). A SM consists of eight streaming processor (SP) cores. 
The Tesla SM uses a new processor architecture called single-instruction, multiple-thread (SIMT). The SIMT unit
creates, manages and executes up to $768$ concurrent threads in hardware with zero scheduling overhead. The SM also implements 
barrier synchronization intrinsic with a single instruction. The fast barrier synchronization, together with 
lightweight thread creation and zero-overhead thread
scheduling support very fine-grained parallelism allowing thousands and even millions of threads to be invoked in kernel calls to achieve
highly scalable parallel programming.

High performance supercomputing has been important in modern astrophysical research since it became available. 
Simulations allow astronomers to perform ``experiments" on astronomical objects, collide stars, 
galaxies, or model the entire visible Universe; all situations are clearly impossible to recreate in a terrestrial laboratory. 
Studying the formation of stars, black holes and galaxies in the Universe is particularly challenging computationally. Their formation
involves the nonlinear interplay of a range of physical processes including gravity, turbulence, magnetic field, shocks, radiation, chemistry, etc. 
Those questions motivated the astrophysical community to develop robust and efficient fluid codes with all the relevant physics.

Studies involving astrophysical fluid dynamics in general are benefitting tremendously from using spatial and temporal adaptive mesh refinement (AMR). This is
especially so in the studies of structure formation. For example, the radius of a star is 8 orders of magnitude smaller than the size of a molecular cloud. 
A uniform grid code is hopeless. On the other hand, the AMR technique has been demonstrated to work
well in resolving the large dynamical range involved in those problems (e.g. Abel et al. 2002; Wang \& Abel 2009). 

The mapping of computational fluid algorithms to GPU however is still at an early stage of development. Harris et al. (2003)
performed cloud simulations using Stam's method (Stam 1999). This method is also used by Liu et al. (2004) for 3D flow calculations. 
Using finite difference methods, Brandvik \& Pulla (2008) solved uniform grid 3D Euler equations, Elsen, LeGresley \& Darve (2008)
solved 3D Euler equations on a multi-block meshes and Zink (2008) solved Einstein's equation with uniform grid. 
As far as we are aware of, this work is the first on mapping an adaptive mesh finite volume solver to GPU. 

\section{CUDA}
\label{sec:CUDA}

In 2007, NVIDIA released CUDA for GPU computing as a language extension to C (NVIDIA 2009). 
CUDA makes GPU computing application development
much easier and more efficient than earlier attempts to GPGPU using various shading languages which need to translate the computation to a graphics language.

CUDA's parallelization model is based on abstraction of the GeForce 8-series hardware. It allows programmer to define kernels which can be executed in parallel by
many threads on GPU. Threads are organized into 1D, 2D or 3D thread blocks, where each block is executed on one SM. 
The SM maps the threads inside a block to the streaming processor (SP) cores and each thread executes independently with its own instruction address and register state. 
Synchronization is possible only within a block whereas global synchronization between blocks is impossible.
Blocks are in turn organized into 1D or 2D grid of
blocks. Each thread can access its thread and block indices by two built-in variables threadIdx and blockIdx. 

The SM's SIMT unit creates, manages, schedules and executes threads in groups of $32$ parallel threads called warps. The threads in a warp always execute 
a common instruction at a time, but different warps execute independently. As a result, different warps can execute on different branches.
This is an enormous improvement for branching code compared to previous-generation GPUs as the $32$-thread warps are much
narrower than the SIMD (single-instruction multiple-data) width of prior GPUs.
However, if threads of a warp diverge via a conditional branch, different execution path have to be serialized, increaing
the total number of instructions executed for this warp. So branching inside a warp should still be minimized to achieve good efficiency.

CUDA exposes the hardware memory hierarchy by allowing threads to access data from multiple memory spaces. All threads have access to the same
global memory. Each thread block has a shared memory visible to all threads of the block and with the same lifetime as the block. Each thread has a
private local memory and a set of registers. There are also two additional read-only memories accessible by all threads: the constant and texture memory.

The shared memory is much smaller than global memory, typically $16$ kB, but it is on-chip so it has very high register-level bandwidth. A typical
programming pattern utilizing this fact is to stage data from global memory into shared memory, process the data there and then write the
results back to global memory.

\section{Block-structured adaptive mesh refinement}

In the Berger \& Colella block-structured AMR (Berger \& Colella 1989), a subgrid will be created in regions of its parent grid needing higher resolution. The hierarchy of grids is organized in a tree structure. Each grid is evolved as a separate initial boundary value problem. The whole grid hierarchy is evolved recursively. In this framework, a single grid is a natural unit to be sent to GPU for computing. Since the grids are dynamically created, arbitrary grid dimensions can arise in real applications. Thus \emph{the key issue for a parallelization scheme to couple efficiently to AMR is to let it work for arbitrary grid dimensions}. Many parallelization models previously proposed for uniform grid GPU fluid solvers are thus not appropriate to be coupled with AMR.

In this work, we use the implementation of Berger \& Colella AMR in the publicly available hydrodynamics code Enzo (Enzo 2009). 
Previously we have implemented our own hydrodynamics and magnetodynamics code using the Enzo AMR framework (Wang et al. 2008; Wang \& Abel 2009).
Often, the hydrodynamics/magnetohydrodynamics solver is the computationally dominant part (typically more than 10 times 
expensive than the AMR part in single core run). So in this work we consider the mapping of hydrodynamics solver onto GPU while leaving the creation and refinement of the grid
hierarchy on CPU. 

\section{Fluid solver}
\label{sec:solver}

The flows involved in star and galaxy formation are highly supersonic and the Reynolds number in those flows are very large (Spitzer 1978).
Thus we are interested in solving the equations of compressible inviscid hydrodynamics. However, the parallelization scheme we implemented
can be applied to any hyperbolic conservation laws. We will demonstrate this explicitly in section \ref{sec:MHD} by implementing a AMR 
magnetohydrodynamics (MHD) solver on GPU.

The equations of compressible inviscid hydrodynamics can be written in the form of conservation laws,
\begin{equation}
 \frac{\partial{U}}{\partial{t}} +
 \frac{\partial{F^x}}{\partial{x}} + \frac{\partial{F^y}}{\partial{y}} + \frac{\partial{F^z}}{\partial{z}}= 0, \label{hydro}
\end{equation}
The conserved variable $U$ is given by
\begin{equation}
 U = (\rho, \rho v_x, \rho v_y, \rho v_z, \rho E)^{T},
\end{equation} 
where $\rho$ is density, $v_i$ are the three components of velocity for $i={x,y,z}$, $E=v^2/2 + e$ is the total energy and $e$ is the internal energy.

The fluxes are given by
\begin{eqnarray}
 F^x &=& \left(\rho v_x, \rho v_x^2+p, \rho v_yv_x, \rho v_zv_x, \rho ({v^2\over2} + h)v_x\right)^{T},\\
 F^y &=& \left(\rho v_y, \rho v_xv_y, \rho v_y^2+p, \rho v_zv_y, \rho ({v^2\over2} + h)v_y\right)^{T}, \\
 F^z &=& \left(\rho v_z, \rho v_xv_z, \rho v_yv_z, \rho v_z^2+p, \rho ({v^2\over2} + h)v_z\right)^{T},
\end{eqnarray}
where $h=e+p/\rho$ is the enthalpy. In this work we assume the ideal gas equation of state:
\begin{equation}
p=(\gamma-1)\rho e,
\end{equation}
where $\gamma$ is the adiabatic index.

Many numerical schemes have been developed to solve hyperbolic conservation laws of the form (\ref{hydro}). We are interested in a class
of finite volume methods called high-resolution shock-capturing (HRSC) schemes developed since the mid-1980s. Those schemes are designed
to capture the correct shock speed even with very low resolutions (see e.g. LeVeque 2002 for a comprehensive review). This property makes
it ideal for adaptive mesh fluid simulations where shocks outside the refined regions may not be well resolved. HRSC schemes represent the most popular scheme in
modern astrophysical codes (e.g. Stone et al. 2008; Mignone et al. 2007; Wang et al. 2008). Thus we will focus on mapping the HRSC schemes onto GPU. 

We use the method of lines (MOL) to discretize the system (\ref{hydro}) spatially,
\begin{eqnarray}
   {dU_{i,j,k}\over dt} &&= 
   - \frac{F^x_{i+1/2,j,k} -
     F^x_{i-1/2,j,k}}{\Delta x} 
   - \frac{F^y_{i,j+1/2,k} -
     F^y_{i,j-1/2,k}}{\Delta y} 
   - \frac{F^z_{i,j,k+1/2} - F^z_{i,j,k-1/2}}{\Delta z}\cr
   &&\equiv L(U), 
   \label{ode}
\end{eqnarray}
where $i, j, k$ refers to the discrete cell index in $x, y, z$ directions, respectively. $F^{x}_{i \pm 1/2,j,k}$, $F^{y}_{i,j\pm
1/2,k}$ and $F^{z}_{i,j,k\pm 1/2}$ are the fluxes at the
cell interface.

As discussed by Shu \& Osher (1988), if one uses a high order scheme to reconstruct flux spatially, one must also use
the appropriate multi-level total variation diminishing (TVD) Runge-Kutta schemes to
integrate the ODE system (\ref{ode}). So instead of the forward Euler time integration
in the original Berger-Collela AMR (Berger \& Colella 1989), we have implemented 
the second order TVD Runge-Kutta scheme:
\begin{eqnarray}
U^{(1)}  &=&  U^n + \Delta{t} L(U^n), \cr
       U^{n+1}&=&{1\over2}U^n+{1\over2}U^{(1)}+{1\over2}\Delta tL(U^{(1)}), \label{rk2}
\end{eqnarray}
See Wang et al. (2008) for further details of the implementation of Runge-Kutta scheme in an AMR framework.

Generally speaking, there are two classes of spatial reconstruction schemes (LeVeque 2002). 
One is reconstructing the unknown variables at the cell interfaces and then use exact or approximate 
Riemann solver to compute the fluxes. Another is direct flux reconstruction, in which we 
reconstruct the flux directly using the fluxes at cell center. We will adopt the first class in this work. But our framework 
can also be applied directly to the second class.

\begin{figure}
\centering
\includegraphics[width=2.2in,height=0.6in]{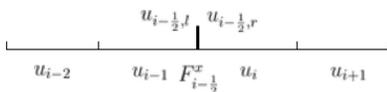}
\caption{The stencil for the calculation of flux $F^x_{i-{1\over2}}$ by equations (\ref{ul}) to (\ref{flux}).} \label{grid1D}
\end{figure}

For example, to calculate the flux $F^x_{i-{1\over2}}$ at cell interface $x_{i-{1\over2}}$ (see Fig.\ref{grid1D}), we need to reconstruct
the left and right states $u_{i-{1\over2},l}$ and $u_{i-{1\over2},r}$. In this work, we use the piecewise linear method (PLM) (van Leer 1979) for reconstruction, 
in which the values of primitive variable $u$ at $x_{i-2},x_{i-1},x_i,x_{i+1}$ are needed. The formula for PLM reconstruction is
\begin{eqnarray}
u_{i-{1\over2},l}&=&u_{i-1}+{\rm minmod}\left((u_{i-1}-u_{i-2})\theta, (u_i-u_{i-1})\theta, (u_i-u_{i-2})/2\right), \\ \label{ul}
u_{i-{1\over2},r}&=&u_{i}+{\rm minmod}\left((u_{i}-u_{i+1})\theta, (u_{i-1}-u_i)\theta, (u_{i-1}-u_{i+1})/2\right),
\end{eqnarray}
where $0\le\theta\le2$ and the minmod function is given as
\begin{equation}
{\rm minmod}(a,b,c)=0.25({\rm sign}(a)+{\rm sign}(b))|{\rm sign}(a)+{\rm sign}(c)|{\rm min}(|a|,|b|,|c|)
\end{equation}

With those two states, we use the Harten-Lax-van Leer (HLL) approximate Riemann
solver (Harten et al. 1983) to calculate the flux $F^x_{i-{1\over2}}$ by,
\begin{equation}
F^x_{i-{1\over2}} = 0.5(F^x(u_{i-{1\over2},l}) + F^x(u_{i-{1\over2},r})-\lambda_0( u_{i-{1\over2},r}-u_{i-{1\over2},l})), \label{flux}
\end{equation}
where $\lambda_0={\rm max}\left({\rm max}(0,\lambda^+_l,\lambda^+_r), {\rm max}(0, -\lambda^-_l,-\lambda^-_r)\right)$ and $\lambda^{\pm}_{i} = v_{x,i}\pm c_{s,i}$ for $i=l,r$.

The important property of this MOL approach is that the cells required for flux calculation in any direction are all along that direction. As we will see below, this property is crucial
for a efficient parallelization scheme in 2D and 3D.

\section{Mapping fluid solver to GPU}

\subsection{One dimensional case}

In this section, we discuss a parallelization scheme for the 1D case. In the next section we will extend this scheme to 2D and 3D. 

First, we note that one drawback of putting all the steps of the 2nd order Runge-Kutta scheme (\ref{rk2}) to GPU is that it needs to send both the old
value and intermediate value to GPU in the second step. From our experience, data transfer between CPU and GPU is quite expensive so it needs to be minimized.
This would also double the GPU memory usage. However, the 2nd order Runge-Kutta scheme can also be written as
\begin{eqnarray}
U^{(1)}  &=&  U^n + \Delta{t} L(U^n), \cr
U^{(2)} &=& U^{(1)}+\Delta tL(U^{(1)}),\cr
U^{n+1}&=&{1\over2}U^n+{1\over2}U^{(2)}.
\end{eqnarray}
Thus, if we only put the first and second step to GPU, which is the computationally intensive part, and leave the third part on CPU, which is 
just the addition of two vectors, we may still get large speed-up. This will be the strategy we take.

Since the first and second steps have exactly the same form, we only need to write one routine computing $U^{(1)}  =  U^n + \Delta{t} L(U^n)$ and 
call it twice. The pseudocode for this routine reads:
\begin{tabbing}
\ \ \ \ \ \ \ \ \=\\
\>allocate memory for primitives and fluxes on GPU\\
\>copy primitives to GPU\\
\>call kernel for flux computation\\
\>call kernel for $L(U)$ computation\\
\>call kernel for time update $U^{(1)}  =  U^n + \Delta{t} L(U^n)$\\
\>copy $U^{(1)}$ back to CPU\\
\>free memory on GPU\\
\end{tabbing}

Since the computation of flux vector $L(U)$ and time update use only local information and involve only one read and one computation, their implementations 
are simple: just launch one thread for every active cell.

\begin{figure}
\centering
\includegraphics[width=4in,height=1in]{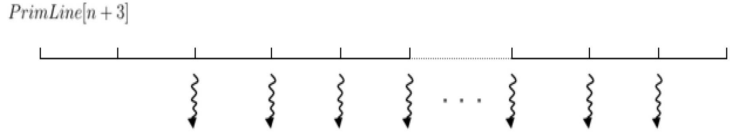}
\caption{The parallelization scheme inside a single block of size $n$. Every thread, except the first and last, loads the data in the cell to the right of it.
The first thread loads in two more cells to the left of it while the last thread loads in one more cell to the right of it.} \label{gridThread}
\end{figure}

The most computational intensive and tricky part is the flux computation, which happens at cell interfaces and needs a different treatment compared to the other
two kernels.
Our basic scheme is that every flux is calculated by a single thread (see Fig. \ref{gridThread}). So for a problem with $N$ grid cells (including the ghost cells),
the number of threads should be $N-3$. We take the number of threads per block to be a fixed number $n=64$. Thus the number of blocks to be 
launched is ${\rm int}(N/64)+1$. The final block 
will have some threads that do not correspond to flux computation, we add a conditional statement in the kernel to 
just let those threads do nothing. Since this affect only the last one or two warps, this remains efficient.

For example, in the $i$th block, we want to calculate $n$ fluxes at cell interfaces
$\{x_{k-{1/2}}, x_{k+{1/2}}, ...,x_{k+n-{3/2}}\}$ with $k=(i-1)n+2$. According to Eqs. (\ref{ul}) to (\ref{flux}), every thread needs 4 cells around it to 
calculate the flux. So in total, $n+3$ cells $\{x_{k-2}, x_{k-1}, ...,x_{k+n}\}$ are needed for the calculation.

As discussed in section \ref{sec:CUDA}, the memory bandwidth is much higher on the shared memory. So in the computation of a block, we will
first load the $n+3$ cells from global memory to a shared memory array PrimLine[n+3]. After the computation is finished, we will then write the
flux back to global memory. As shown in Fig.\ref{gridThread}, our memory loading scheme goes as follows: thread $i$ will read in primitive variables in cell $x_{k+i+2}$. To read in all the
necessary data, the zeroth thread will then need to read in two more cells $x_{k}$ and $x_{k+1}$ and the (n-1)th thread will read in one more cell $x_{k+n+2}$.

The kernel code for this flux computation are:
\\

\begin{tabbing}
\#define CUDA\_BLOCK\_SIZE 64\\
\#define NEQ\_HYDRO 5\\
...\\
\\
 // get thread and block index\\
 const long tx = threadIdx.x;\\
 const long bx = blockIdx.x;\\
\\
 int igrid = tx + bx*CUDA\_BLOCK\_SIZE; // array index in the data field\\
 int idx\_prim, idx\_prim1;\\
\\
 \_\_shared\_\_ float PrimLine[NEQ\_HYDRO*(CUDA\_BLOCK\_SIZE+3)]; // input primitive variable\\
 \_\_shared\_\_ float FluxLine[NEQ\_HYDRO*CUDA\_BLOCK\_SIZE]; // output flux\\
\\
 if \= (igrid $>=$ 2 \&\& igrid $<=$ size - 2) \{ // only do flux computation for active cells\\
\\
\>   // load data from device to shared.\\
 \>  idx\_prim1 = (tx+2)*NEQ\_HYDRO;\\
  \> PrimLine[idx\_prim1++] = Rho[igrid];\\
  \> PrimLine[idx\_prim1++] = Eint[igrid];\\
  \> PrimLine[idx\_prim1++] = Vx[igrid];\\
  \> PrimLine[idx\_prim1++] = Vy[igrid];\\
  \> PrimLine[idx\_prim1] = Vz[igrid];\\

   \>// if the first, load in two more cells for boundary condition\\
   \>if (tx == 0 $||$ igrid == 2) \{\\
     \>\ \ for (int i = -2; i $<=$-1; i++) \{\\
	\> \ \ \ \ idx\_prim  = igrid + i;\\
       \>\ \ \ \ idx\_prim1 = (i+tx+2)*NEQ\_HYDRO;\\
       \>\ \ \ \ PrimLine[idx\_prim1++] = Rho[idx\_prim];\\
	\>\ \ \ \ PrimLine[idx\_prim1++] = Eint[idx\_prim];\\
       \>\ \ \ \ PrimLine[idx\_prim1++] = Vx[idx\_prim];\\
	\>\ \ \ \ PrimLine[idx\_prim1++] = Vy[idx\_prim];\\
       \>\ \ \ \ PrimLine[idx\_prim1] = Vz[idx\_prim];\\
     \> \ \ \}\\
   \>\}\\
   \\
       \>// if the last, load in one more cell for boundary condition\\
   \>if (tx == CUDA\_BLOCK\_SIZE - 1 $||$ igrid == size - 2) \{\\
     \>\ \ idx\_prim  = igrid + 1;\\
     \>\ \ idx\_prim1 = (tx+3)*NEQ\_HYDRO;\\
     \>\ \ PrimLine[idx\_prim1++] = Rho[idx\_prim];\\
     \>\ \ PrimLine[idx\_prim1++] = Eint[idx\_prim];\\
     \>\ \ PrimLine[idx\_prim1++] = Vx[idx\_prim];\\
     \>\ \ PrimLine[idx\_prim1++] = Vy[idx\_prim];\\
     \>\ \ PrimLine[idx\_prim1] = Vz[idx\_prim];\\
  \>\}\\
  \}\\
  \\
    // synchronize to ensure all the data are loaded\\
 \_\_syncthreads();\\
 \\
   if (igrid $>=$ 2 \&\& igrid $<=$ size - 2) \{\\
   \>// the main computation: calculating the flux at tx\\
   \>LLF\_PLM(PrimLine, FluxLine, tx);\\
\\
   \>// copy 1D Flux back to Flux\\
   \>idx\_prim1 = tx*NEQ\_HYDRO;\\
   \>FluxD[igrid] = FluxLine[idx\_prim1++];\\
   \>FluxS1[igrid] = FluxLine[idx\_prim1++];\\
   \>FluxS2[igrid] = FluxLine[idx\_prim1++];\\
   \>FluxS3[igrid] = FluxLine[idx\_prim1++];\\
   \>FluxTau[igrid] = FluxLine[idx\_prim1  ];\\
 \}
\end{tabbing}

\subsection{Two and three dimensional case}

In higher dimension, the main trick is that in the MOL, 2D and 3D flux computation can be reduced to 1D problems and thus the parallelization 
scheme discussed above for 1D problem can be applied directly to 2D and 3D problems.

\begin{figure}
\centering
\includegraphics[width=4.7in,height=2.5in]{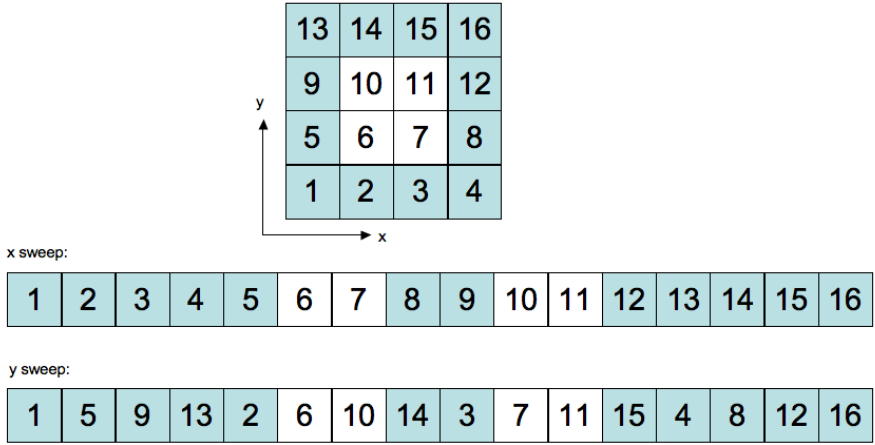}
\caption{This figure illustrates how the 2D grid with active (white) and ghost (shaded) cells are mapped to 1D grids in $x$ and $y$ sweeps.} \label{sweep}
\end{figure}

For example, in 2D application of MOL, one sweeps through a 1D grid
of lines extending in the other direction, calculating the fluxes on every line in every sweep. To compute flux in all two directions, one does two sweeps. The trick is that in every sweep, we conceptually regard the 2D grid to be a 
large 1D grid, including the ghost cells, which is continuous in the direction we are sweeping. Thus, when we apply thread decomposition to the grid, 
we send the large 1D array to it and use exactly the same scheme discussed in previous section for 1D problems. This is illustrated in Fig. \ref{sweep} for a 2D problem, where we show the original 2D grid and the 1D grid sent to the kernel for $x$ and $y$
sweeps. The 3D case is exactly the same. The additional cost of this method is that we will also calculate fluxes for all the ghost cells since now
we regard them as normal cells so some calculation is wasted. However, this is only a small cost. But the gain is that the parallelization scheme
now works for any grid dimensions. One only need to use two ``if" statements to handle the first and last points in the grid, which will lead to 
execution branching only in the first and last two warps. Note that in $y$ sweeping, when reading 
from global to shared memory, the reading is from non-continuous locations. Thus reading is non-coalesced in those cases. We have also experimented 
reshuffling the large 3D array in CPU so that the reading can be coalesced but we found this leads to lower performance because of the reshuffling cost on CPU
and additional data transfer.

\section{Results}

\begin{table}
\label{Quadro}
\caption{Technical specifications of NVIDIA's Quadro FX 5600 graphics card.}
\centering
\begin{tabular}{c c}
\\
\hline
Number of streaming multiprocessors (SM) & 16 \\
Number of streaming processors per SM & 8\\
Warp size & 32\\
Parallel data cache & 16 kB\\
Number of banks in parallel data cache & 16\\
Number of 32-bit registers per SM & 8192 \\
Clock frequency of each SM & 1.35 GHz\\
Frame buffer memory type & GDDR3\\
Frame buffer interface width & 384 bits\\
Frame buffer size & 1.5 GB\\
Constants memory size & 64 kB\\
Clock frequency of the board & 800 MHz\\
Host bus interface & PCI Express\\
\hline
\end{tabular}
\end{table}

All the problems in this work are run on a Quadro FX 5600 card. For reader's convenience, the technical specifications of Quadro FX 5600 card
are listed in Table 1. The corresponding CPU comparison cases are run on a single $3$ GHz core.

\subsection{1D Sod problem}
\label{sec:sod}

\begin{figure}
\centering
\includegraphics[width=4.in,height=2.5in]{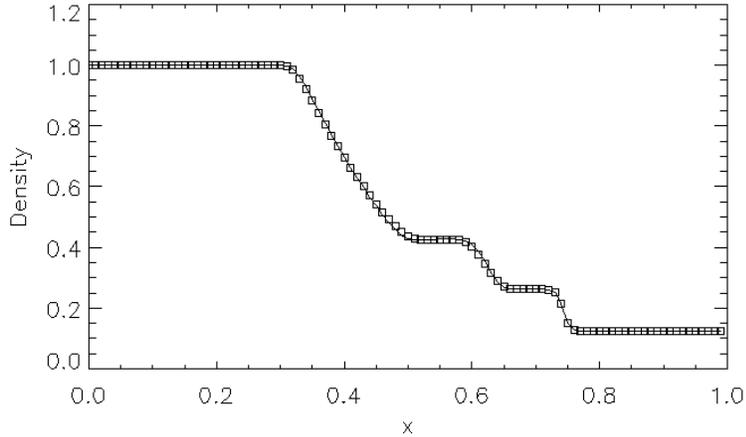}
\caption{Density profile at $t=0.14$ for the Sod test problem with $N=100$. Solid line is the CPU solution and squares are the GPU solution.} \label{sod}
\end{figure}

To evaluate the accuracy of the GPU solution, we first run a 1D Sod shock tube problem (Sod 1978) with both
the CPU code and GPU code. The initial condition for this problem is two uniform states separated at $x=0.5$. 
The left and right density and pressure are $\rho_l = 1, p_l=1$ and $\rho_r=0.125,p_r=0.1$. The initial
velocity is zero everywhere and the adiabatic index is $\gamma=1.4$. We use $100$ grid points for this test. The result is shown in Fig.\ref{sod}. 
It can be seen that the GPU solution agrees with the CPU solution very well. This validates our GPU implementation of fluid solver.

\subsection{3D Sedov-Taylor blast wave}
\label{sec:sedov}

\begin{figure}
\centering
\includegraphics[width=4.in,height=3in]{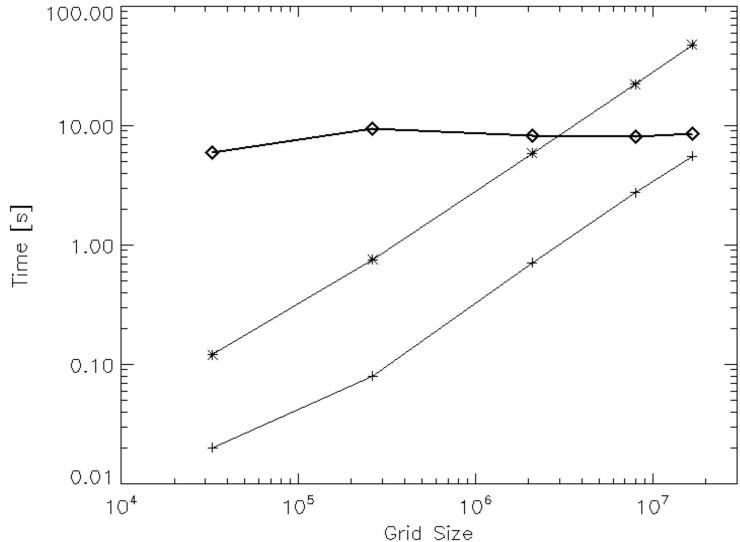}
\caption{Running time of a single timestep for CPU code (asterisk) and GPU code (plus) for 3D uniform grid Sedov-Taylor blast wave problem with grid sizes
$N=32^3,64^3,128^3,200^3,256^3$. 
The diamonds show the ratio of the CPU and GPU running time.} \label{SedovTime}
\end{figure}

Next, we study the performance of GPU code in 3D using the Sedov-Taylor blast wave problem (Sedov 1959). This section will use a uniform grid. An AMR example is given in 
the next section. 

To set up the problem, we deposit a total energy $E=1$ into a spherical region of radius $r_0=0.1$ (the ``explosion region") 
at the center of the simulation box which has length $1$. The 
pressure inside the explosion region can then be calculated as $p=3(\gamma-1)E/(4\pi r_0^3)$, where $\gamma=1.4$ for this problem. 
The initial density is uniform throughout the box with $\rho=1$ and the pressure is set to a small value $p=10^{-5}$ outside
the explosion region. 

Fig.~\ref{SedovTime} shows the performance comparison of the CPU and GPU code for uniform grid sizes
$N=32^3, 64^3, 128^3,$ $200^3, 256^3$. It can be seen that we get fairly uniform $8-10$ times speed-up for two orders of magnitude difference
in grid sizes. This is very encouraging as a wide variety of grid sizes can arise in real AMR applications and a uniform speed-up for 
a large range of grid sizes is a necessary condition for good performance in AMR applications.

\subsection{3D cloud disruption with AMR}

In this section we show a case of cloud disruption by blast wave, demonstrating that our implementation
also lead to good speed-up on AMR applications. We put a uniform density cloud with radius $0.01$ at distance $0.15$ away from the center of the blast wave. 
The cloud is $10$ times denser than the medium and is in pressure equilibrium with the medium. The topgrid has resolution $64^3$ and we use six levels of refinement to
resolve the cloud disruption, which correspond to an effective uniform resolution $4096^3$.

For this problem, the running time of a single timestep in GPU case is $\sim8$ times faster than the CPU case. This is consistent with the behavior we saw in last section for uniform grid 
tests with various grid sizes.

\begin{figure}
\centering
\includegraphics[width=5in,height=2.5in]{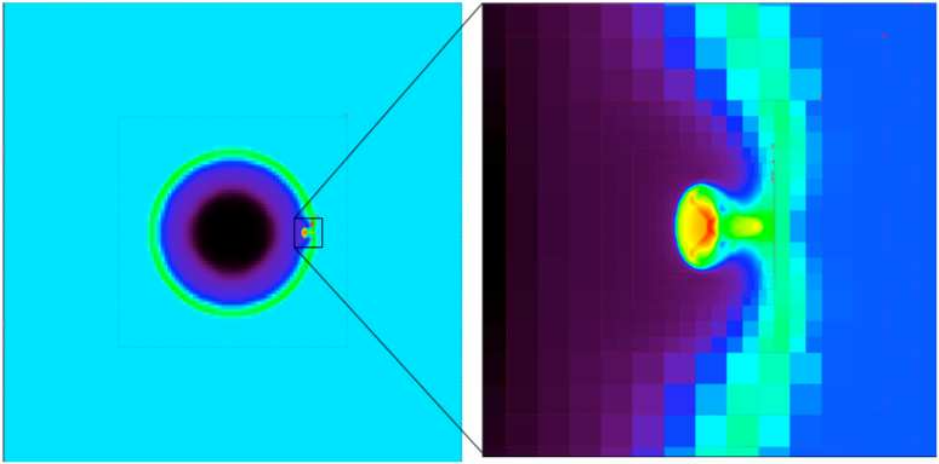}
\caption{Density slice showing the disruption of a cloud by a Sedov-Taylor blast wave. The topgrid resolution is $64^3$. Six levels of refinement is used to 
resolve the cloud disruption. The dotted lines show the boundary of subgrids.} \label{cloud}
\end{figure}

\section{Magnetohydrodynamics}
\label{sec:MHD}

Magnetic field has been known to play a very important role in the formation of stars (Shu et al. 1987) and its role in galaxy formation has also been
investigated recently (Wang \& Abel 2009). Thus for astrophysical applications, a MHD solver is of great interest. Our parallelization scheme discussed above
can be easily extended to the case of MHD. In this section, after discussing the MHD equations, we will show some results of this GPU MHD solver.

\subsection{Equations}

MHD equations can also be written in the form of conservation laws. Thus any schemes for hydrodynamics in principle can also be applied 
to MHD. The main numerical problem of solving MHD equations is cleaning up the numerically generated magnetic monopoles as magnetic field is
divergence-free physically. Various schemes have been proposed for this purpose and there is still no universal agreement which
scheme is the best (see Toth 2000 for a comparison of various schemes). In this work, we will adopt the so-called hyperbolic clean approach 
(Dedner et al. 2002), which nicely fit in our developed framework. This may not be the best scheme for all applications. But other schemes, like the projection scheme, requires
the solution of a Poisson equation. Thus mapping them to GPU requires additional work on sparse matrix solvers. 

Following Dedner et al. (2002), we consider the generalized Lagrange multiplier 
(GLM) formulation of the MHD equations, which can be written in the conservative form (\ref{hydro}) with the conserved variables given by
\begin{equation}
 U = (\rho, \rho v_x, \rho v_y, \rho v_z, \rho E+B^2/2, B_x, B_y, B_z, \psi)^{T},
\end{equation} 
where $B_i$ with $i={x,y,z}$ are the three components of magnetic fields and 
$\psi$ is the additional scalar field introduced in the GLM formulation for the divergence cleaning.

And the fluxes are given by
\begin{eqnarray}
 F^x &=& (\rho v_x, \rho v_x^2+p+B^2/2-B_x^2, \rho v_yv_x-B_yB_x, \cr
 && \rho v_zv_x-B_zB_x, \rho ({v^2\over2} + h)v_x+B^2v_x-B_xB\cdot v, \cr
&& \psi, v_xB_y-v_yB_x, -v_zB_x+v_xB_z, c_h^2B_x)^T,\\
 F^y &=& (\rho v_y, \rho v_xv_y-B_xB_y, \rho v_y^2+p+B^2/2-B_y^2, \cr
 && \rho v_zv_y-B_zB_y, \rho ({v^2\over2} + h)v_y+B^2v_y-B_yB\cdot v, \cr
 && v_yB_z-v_zB_y,\psi,-v_xB_y+v_yB_x, c_h^2B_y)^{T}, \\
 F^z &=& (\rho v_z, \rho v_xv_z-B_xB_z, \rho v_yv_z-B_yB_z, \rho v_z^2+p+B^2/2-B_z^2, \cr
 && \rho ({v^2\over2} + h)v_z+B^2v_z-B_zB\cdot v, \cr
    &&  -v_yB_z+v_zB_y, v_zB_x-v_xB_z,\psi, c_h^2B_z)^{T},
\end{eqnarray}
where $c_h$ is a constant controlling the propagation speed and damping rate of $\nabla\cdot B$ (Dedner et al. 2002). 

Solving the GLM-MHD system in our framework is straightforward. All we need to do is to 
add the additional primitive variables to our hydrodynamics solver.

\subsection{1D Brio-Wu problem}

\begin{figure}
\centering
\includegraphics[width=4.in,height=2.5in]{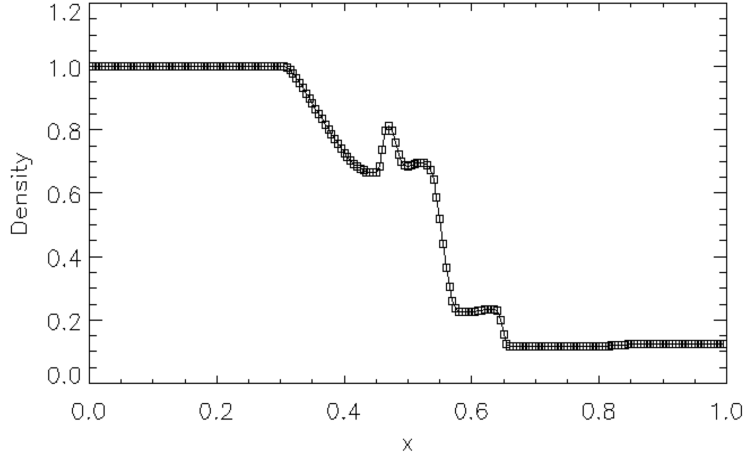}
\caption{Density profile at $t=0.1$ for the Brio-Wu test problem with $N=200$. Solid line is the CPU solution and squares are the GPU solution.} \label{briowu}
\end{figure}

First, to evaluate the accuracy of the GPU solution, we run a 1D Brio-Wu shock tube problem (Brio \& Wu 1988) with both
the CPU code and GPU code. The setup of Brio-Wu problem is similar to the Sod problem discussed in section \ref{sec:sod},
with two uniform states separated at $x=0.5$.
But here there is non-zero magnetic field in the initial condition.  
The left and right states are $\rho_L = 1, p_L = 1, v_{xL} = 0, v_{yL} = 0, B_{xL} = 0.75, B_{yL} = 1$ and $\rho_R = 0.125, p_L = 0.1, v_{xL} = 0, v_{yL} = 0, B_{xL} = 0.75, B_{yR} = -1$. The adiabatic index is taken to be $\Gamma=2$. We use $200$ grid points for this test. The result is shown in Fig.\ref{briowu}. 
It can be seen that the GPU solution agrees with the CPU solution very well. This validates our GPU implementation of MHD solver.

\subsection{Sedov-Taylor Blast Wave}

\begin{figure}
\centering
\includegraphics[width=4.in,height=3in]{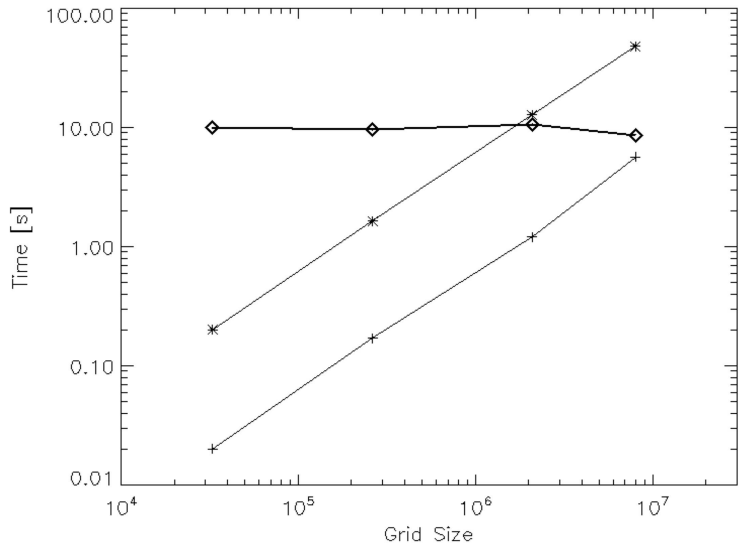}
\caption{Running time of MHD code on a single timestep for CPU code (asterisk) and GPU code (plus) for 3D uniform grid Sedov-Taylor blast wave problem with grid sizes
$N=32^3,64^3,128^3,200^3$. 
The diamonds are the ratio of the two running times.} \label{MHDSedovTime}
\end{figure}

To compare the performance of the MHD solver to the Hydrodynamics solver, we apply the MHD solver to the Sedov-Taylor blast wave problem discussed in section \ref{sec:sedov}.
Fig.~\ref{MHDSedovTime} shows the performance comparison of the CPU and GPU MHD solvers for uniform grid sizes
$N=32^3,64^3,128^3,200^3$. Note that for the MHD case, the memory requirement for a $256^3$ run is $1.81$ GB (three--primitives, flux and $L(U)$--copies of the nine 
MHD fields in a grid size $256^3$). Thus it cannot fit in the $1.5$ GB memory of a Quadro FX 5600 card.

From Fig.\ref{MHDSedovTime} we can see that in the MHD case we also get $8-10$ speedup for a large range of grid sizes. This demonstrates the efficiency of our scheme 
as applied to more complicated physical problems.

\subsection{MHD turbulence}

\begin{figure}
\centering
\includegraphics[width=4.in,height=4in]{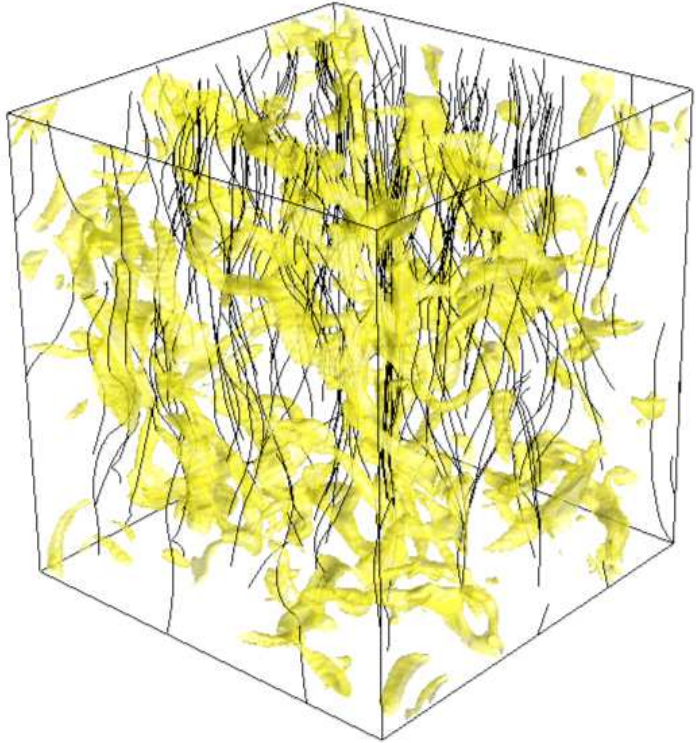}
\caption{MHD turbulence simulation: density iso-surfaces with magnetic field lines at one dynamical time.} \label{turbulence}
\end{figure}

\begin{figure}
\centering
\includegraphics[width=3.in,height=2in]{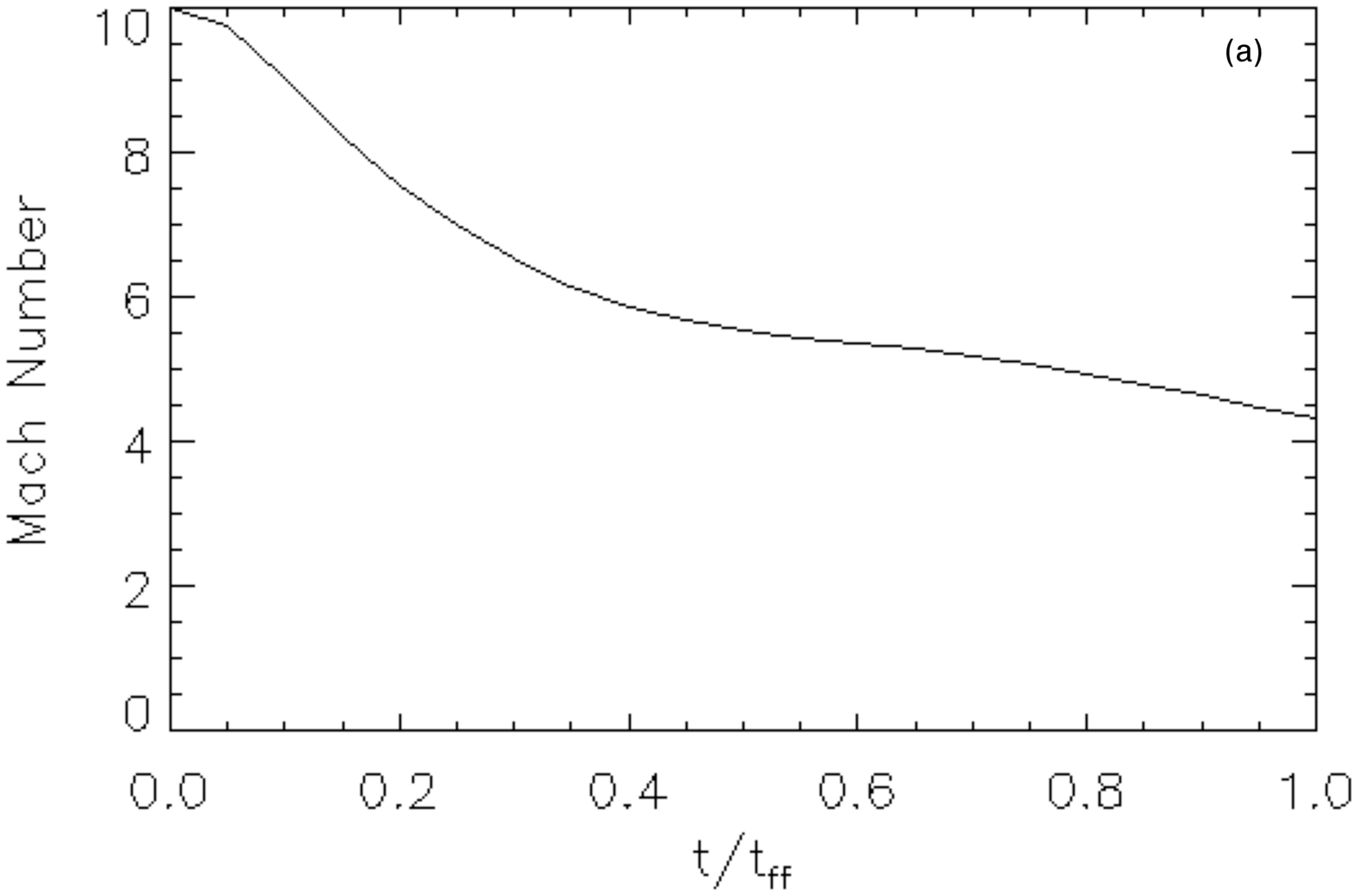}
\includegraphics[width=3.in,height=2in]{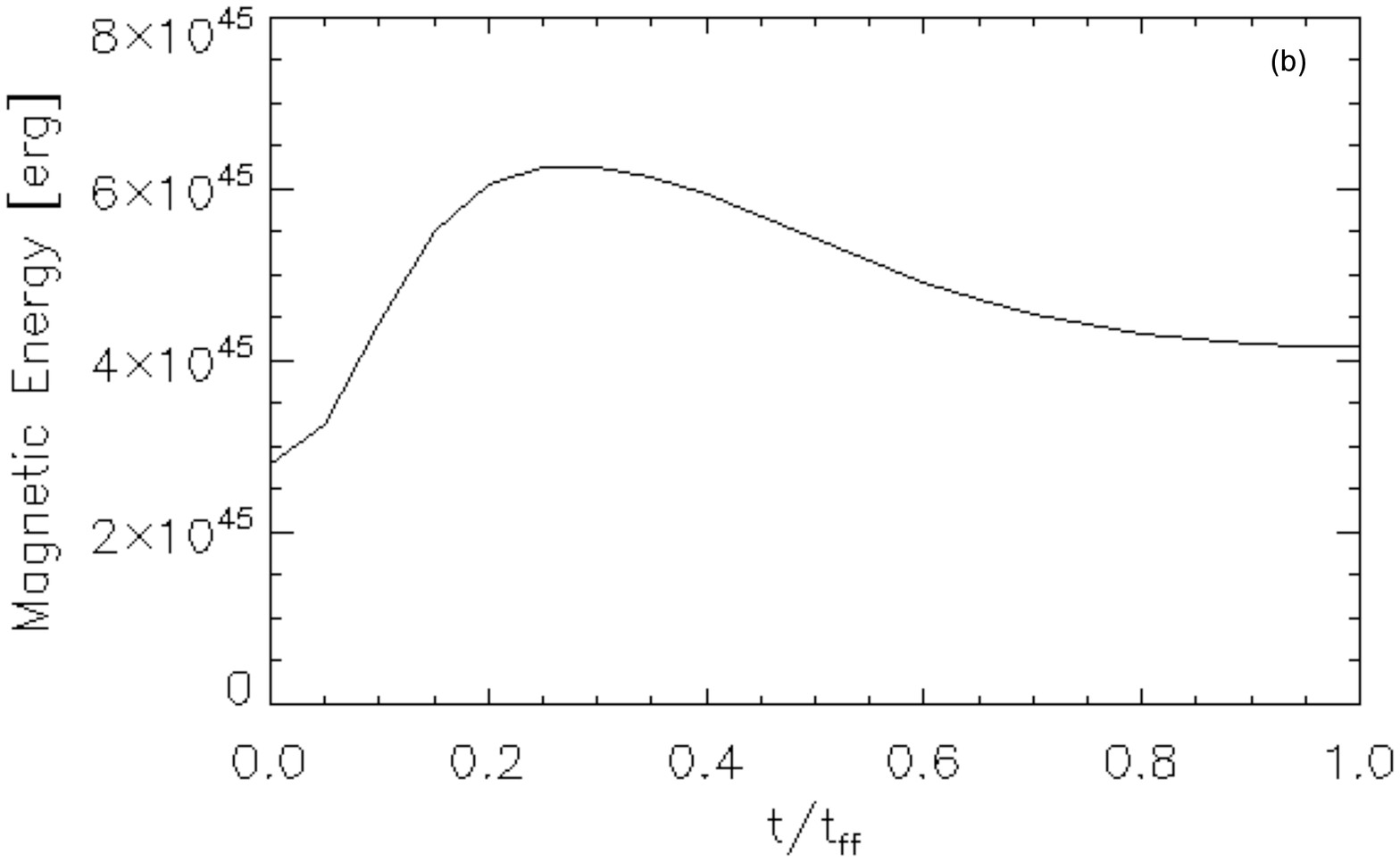}
\caption{Time evolution of averaged Mach number (a) and total magnetic energy (b).} \label{mach}
\end{figure}

Finally, we present the results of an application of our GPU MHD solver to a problem that is of great astrophysical interest related to star formation: the decay of MHD turbulence (Stone et al. 1998; Mac Low et al. 1998).
Following Stone et al. (1998), we set  initial density to be uniform. An isotropic turbulent velocity with Burger-like power spectrum $P(k)\propto k^{-2}$ is imposed. The initial Mach number 
is $\mathcal{M}=10$. We use an isothermal equation of state by setting $\gamma=1.001$. The resolution is taken to be uniform $128^3$. 
Fig.~\ref{turbulence} shows the results of density field after one dynamical time. The code takes about $10$ minutes to evolve to this point. Fig.~\ref{mach}
shows the time evolution of the average Mach number and total magnetic energy. Consistent with the findings in Stone et al. (1998), a significant
fraction of the kinetic energy decays in one dynamical time, and the presence
of magnetic field cannot delay it. Those results are not new. However, the 
much shorter running time of the GPU code makes it possible to do many runs in a reasonable time to build up statistics,
which is crucial for some aspects of star formation such as the core mass function. 
Currently, this has only been possible in 2D (e.g. Basu \& Ciolek 2009). 

\section{GPU cluster computing}

\begin{figure}
\centering
\includegraphics[width=3.in,height=2.5in]{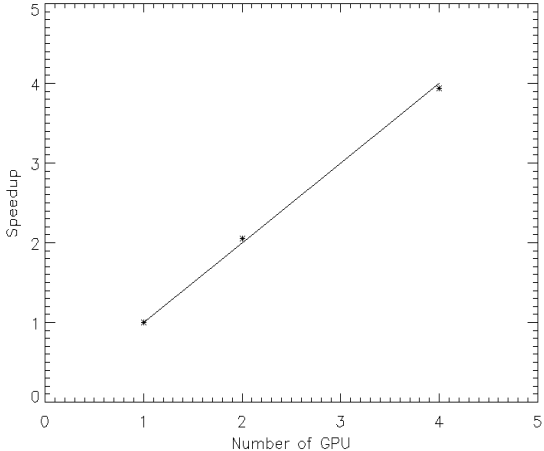}
\caption{Speedup of the MPI-CUDA hybrid code as a function of the number of GPUs for uniform $128^3$ resolution Sedov-Taylor blast wave simulations.} \label{speed}
\end{figure}

To be competitive with current state of the art simulations which are usually run on large CPU clusters, a GPU implementation should also be MPI parallel. A simple setup would be building a small CPU cluster
with each CPU node containing one or a few GPU cards. As a simple experiment, we built a 4 nodes CPU cluster with 4 Mac Pro workstations and a NVIDIA GeForce 8800 GT graphics card on each node.
The CPU nodes communicate with each other using the standard MPI library over Gigabit ethernet.

As discussed above, in our scheme, a grid is the basic unit of GPU computing. This makes it trivial to combine our scheme with the standard MPI parallel scheme of Berger-Collela AMR, which also use
grids as the basic unit for distribution among processors. The code first calls MPI library to distribute grids to different processors. Then each processor sends its grids, one by one, to its GPU for computing.
We have tested the resulting MPI-CUDA hybrid code on up to 4 GPUs. The results of the
Sedov-Taylor blast wave simulations with a uniform $128^3$ resolution using 1, 2 and 4 GPUs are shown in Fig.~\ref{speed}. It shows that 
our code archived very good, close to ideal speedup for up to four GPUs. Thus it seems realistic that large GPU clusters may offer significant cost saving as compared to a CPU clusters giving the same performance the magneto hydrodynamical problems our implementation can study. 

\section{Discussions and future directions}

In this work we described how to map HRSC schemes for hyperbolic conservation laws to GPU using NVIDIA's CUDA. We demonstrated that 
our framework as applied to the equations of inviscid compressible hydrodynamics and MHD can lead to a significant speedup. Specifically,
on a Quadro FX 5600 card, saw approximately a factor of ten speedup compared to a single 3 GHz CPU core. 

An important purpose of our GPU parallel scheme design is achieving good scalability. 
In typical fluid simulations with as many as millions of cells per grid, our parallelization scheme will launch millions of threads at the same time on GPU
to perform the computation. This exceeds the Quadro FX 5600's ability of running $12288$ threads concurrently by more than
two orders of magnitude. Thus we expect that the speedup factor of our parallelization scheme
will increase linearly with the number of SMs on the GPU. This makes our scheme highly scalable to future generation of graphics hardwares.

One important topic we would like to concentrate on in the near future is sparse matrix solvers for Poisson equation. If one can also speed up 
Poisson solvers by a similar factor on GPU, then coupled with the fluid solvers we implemented in this work, a whole range of
astrophysical simulations will be open to processing on the GPU as we can then model astrophysical fluids with self-gravity, viscosity and other non-ideal 
effects. An implementation of Poisson solver will also make it possible to use projection method for divergence cleaning and implicit fluid solvers.

\section*{Acknowledgement}

We would like to thank Sean Treichler for helpful discussions. 
We are also grateful to the SLAC computing group, especially Adeyemi Adesanya, Stuart Marshall, Ken Zhou, for technical support on GPU hardware.

\end{document}